# Privacy Things:

## Systematic Approach to Privacy and Personal Identifiable Information

Sabah Al-Fedaghi
Computer Engineering Department
Kuwait University
Kuwait
sabah.alfedaghi@ku.edu.kw

*Abstract*—Defining privacy and related notions such as Personal Identifiable Information (PII) is a central notion in computer science and other fields. The theoretical, technological, and application aspects of PII require a framework that provides an overview and systematic structure for the discipline's topics. This paper develops a foundation for representing information privacy. It introduces a coherent conceptualization of the privacy senses built upon diagrammatic representation. A new framework is presented based on a flow-based model that includes generic operations performed on PII.

*Keywords—Conceptual model, information privacy, identification, Personal Identifiable Information (PII), identifiers*

## I. INTRODUCTION

Privacy has been developed over the years as an applicable field of study in engineering systems. According to Spiekermann and Cranor [1], "Privacy is a highly relevant issue in systems engineering today. Despite increasing consciousness about the need to consider privacy in technology design, engineers have barely recognized its importance." Privacy engineering is concerned with providing methodologies, tools, and techniques for privacy, and it has materialized as an emerging discipline as enterprises increasingly turn to Internet-based cloud computing. Without privacy engineering incorporated into the design, initiation, implementation, and maintenance of cloud programs, data protection and accessibility standards will become increasingly challenging for agencies to properly control [2]. The 2018 EU General Data Protection Regulation can require organizations to pay a fine (4% of their global annual turnover or €20M, whichever is greater) for the most serious infringements of privacy regulations. "Privacy laws are suddenly a whole lot more costly to ignore" [3].

Nevertheless, a 2017 commissioner report [4] complains that privacy across the various sectors tends to be *quite vague* and is often expressed in a language that makes it difficult to apply. For example, it is protested that Google's privacy policy is too vague for users to control how their information is shared.

The meaning of privacy has been much *disputed* throughout its history in response to wave after wave of new technological capabilities and social configurations. The current round of disputes over privacy fueled by data science has been a cause of despair for many commentators and a death knell for privacy itself for others. [5] (Italics added)

After years of consultation and debate, experts and policy-makers have developed protection principles for privacy that form a shared set of fair information practices and have become the basis of personal data or information privacy laws in much institutional and professional work across the public and private sectors [6].

However, these principles have proved less useful with the rise of data analytics and machine learning. Informational self-determination can hardly be considered a sufficient objective, nor individual control a sufficient mechanism, for protecting privacy in the face of this new class of technologies and attendant threats. [5]

Individual control offers no protection or remedy [7] against techniques such as inference, modern forms of data analysis [8] [9] [10], analysis of social media behavior [11], or cross-referencing of "de-identified" data [12].

According to Jones [13], recognizing the senses in which information can be said to be personal "can form a yardstick by which to evaluate supporting tools, organizing schemes and overall strategies in a practice" of handling Personal Identifiable Information (PII). This paper aims at this objective of *recognizing the senses of PII*. "What is PII? Is it personal?" "Personal information" typically refers to information that uniquely identifies an individual [14]. Waling and Sell [15] include the notion of identifiability in their definition: "Personal information is all the recorded information about an identifiable individual."

The important issue in this context is defining the elementary constituents or fundamental units of privacy. Spiekermann and Cranor [1] use at least 11 terms to name the types of "data" involved in privacy: personal data, personally identifiable data, personal information, identifying data, identifiable personal data, privacy informationidentifying information, personally identifiable information, identity information, and privacy related information. They do not explicitly define these types of data. This is a serious issue because the data are *things* around which privacy revolves.

The theoretical, technological, and application aspects of PII require a framework that provides a general view and a systematic structure for the discipline's topics. This paper uses a diagrammatic language called Flowthing Machines (FM) to develop a framework for a firmer foundation and more coherent structures in privacy.

The FM model used in this paper is a diagrammatic representation of "things that flow." *Things* can refer to a range of items including data, information, and signals. Many scientific fields use diagrams to depict knowledge and to assist in understanding problems. "Today, images are ... considered not merely a means to illustrate and popularize knowledge but rather a genuine component of the discovery, analysis and justification of scientific knowledge" [16]. "It is a quite recent movement among philosophers, logicians, cognitive scientists and computer scientists to focus on different types of representation systems, and much research has been focused on diagrammatic representation systems in particular" [17].

For the sake of a self-contained paper, we briefly review FM, which forms the foundation of the theoretical development in this paper. It involves a diagrammatic language that has been adopted in several applications [18-22]. The review is followed by sections that introduce basic notions that lead to defining of PII. Section 3 explores the notion of a *signal* as a vehicle that carries data, which leads to defining *data* and *information* (section 4), to arrive at the fundamental notion of *identifier* (section 5), thus arriving at privacy concepts and PII. Section 6 defines PII and leads to an examination of the question, What is Privacy? in section 7. The remaining sections analyze types of PII, the nature of PII, trivial PII, and sensitive PII.

## II. FLOWTHING MACHINES (FM)

The notion of *flow* was first propounded by Heraclitus, a pre-Socratic Greek philosopher who declared that "everything flows." Plato explained this as, "Everything changes and nothing remains still," where instead of "flows" he used the word "changes" [23]. Heraclitus of Ephesus (535–475 BCE) was a native of Ephesus, Ionia (near modern Kuşadası, Turkey). He compared existing things to the flow of a river, including the observation that you cannot step twice into the same river. Flow can also be viewed along the line of "process philosophy," "championed most explicitly by Alfred N. Whitehead in his 'philosophy of organism,' worked out during the early decades of the 20th century" [23].

According to Henrich et al. [24], flows can be conceptualized as transformation (e.g., inputs transform into outputs),

> Anybody having encountered the construction process will know that there is a plethora of flows feeding the process. Some flows are easily identified, such as materials flow, whilst others are less obvious, such as tool availability. Some are material while others are non-material, such as flows of information, directives, approvals and the weather. But all are mandatory for the identification and modelling of a sound process.

*Things that flow* in FM refer to the exclusive (i.e., being in one and only one) conceptual movement among six states (stages): transfer, process, create, release, arrive, and accept, as shown in Fig. 1. It may be argued that things (e.g., data) can also exist in a *stored* state, which is not included as a stage of FM, however, because *stored* is not a primary state; data can be stored after being created, hence becoming *stored created data*, or after being processed and becoming *stored processed data*,... Current models of software and hardware do not differentiate between these states of stored data. The machine of Fig. 1 is a generalization of the typical input-process-output model used in many scientific fields.

To exemplify FM, consider flows of a utility such as electricity in a city. In the power station, electricity is *created*, then released and *transferred* through transmission lines to city substations, where it arrives. The substations are safety zones where electricity is *accepted* if it is of the right type voltage; otherwise it is cut off. Electricity is then *processed*, as in the case of creating different voltage values to be sent through different feeders in the power distribution system. After that, electricity is *released* from the distribution substation to be *transferre*d to homes. *Receive* in Fig. 1 refers to a combined stage of *Arrive* and *Accept* for situations or scenarios where arriving things are always accepted.

The FM diagram is analogous to a map of city streets with arrows showing the direction of traffic flows. It is a conceptual description because it omits specific details of\ characteristics of things and spheres. All types of synchronization, logical notions, constraints, timing, ... can be included or superimposed on this conceptual representation, in the same way traffic controls, signals, and speed constraints can be superimposed on a map of city streets.

Each type of flow is distinguished and separated from other flows. No two streams of flow are mixed, analogous to separating lines of electricity and water in blueprints of buildings. An FM representation need not include all the stages; for example, an archiving system might use only the stages Arrive, Accept, and Release. Multiple systems captured by FM can interact with each other by triggering events related to one another in their spheres and stages.

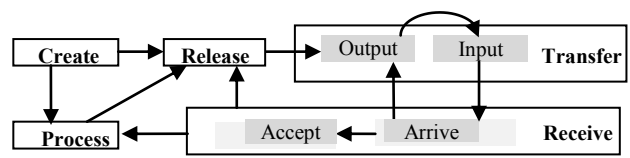

Fig. 1. Flowthing machine.

The fundamental elements of FM are described as follows:
**Things**: A *thing* is defined as *what is being created, released, transferred, arrived, accepted, and processed* while flowing within and between machines. For example, *heat* is a thing because it can be created, processed, ... Similarly, time, space, a contradictory statement, an electron, events, and noise are all things. Mathematical class, members, and numbers are things because they can be created, released, transferred, etc. "Operations" described in verbs such as *generate* are not a thing but another name for the stage *Create*. In FM there are only the "operations" Create, Process Release, Transfer, and Receive (assuming that all arriving things are accepted). Thus, "change" or "sort" is *Process*, "transport" or "send" is

*Transfer*, and a product waiting to be shipped is a *Released* product.

**A machine**, as depicted in Fig. 1, comprises the internal flows (solid arrows) of things along with the stages and transactions among machines.

**Spheres and subspheres** are the environments of the thing, e.g., the stomach is a food-processing machine in the *sphere* of the digestive system. The machines are embedded in a network of spheres in which the processes of flow machines take place. A sphere can be a person, an organ, an entity (e.g., a company, a customer), a location (a laboratory, a waiting room), a communication medium (a channel, a wire). A flow machine is a subsphere that embodies the flow; it itself has no subspheres.

**Triggering** is a transformation (denoted by a dashed arrow) from one flow to another, e.g., a flow of electricity triggers a flow of air. In FM, we do not say, *One element is transformed into another*, but we say *One element is processed to trigger the creation of another*. An element is never changed into a new element; rather, if 1 is a number and 2 is a number, the operation '+' does not transform 1 and 2 into 3, but '+' triggers the *creation* of 3 from input of 1 and 2.

There are many types of flow things, including data, information, money, food, fuel, electrical current, and so forth. We will focus on *information* flow things.

FM is a modeling language. "A model is a systematic representation of an object or event [a thing in FM] in idealized and abstract form… The act of abstracting eliminates certain details to focus on essential factors" [25]. A model provides a vocabulary for discussing certain issues and is thus more like a tool for the scientist than for use in, for instance, practical systems development [26].

We will now introduce basic notions that lead to defining PII. To reach this definition, we explore the notion of a *signal* as a vehicle that carries data, a notion that leads to defining information, to arrive at the fundamental notion of unique identifiers. This provides a way to define privacy and PII.

### III. WHAT IS A SIGNAL?

The flow of things seems to be a fundamental notion in the world. According to NPTEL [27],

> We are all immersed in a sea of *signals*. All of us from the smallest living unit, a cell, to the most complex living organism (humans) are all the time receiving signals and processing them. Survival of any living organism depends on processing the *signals* appropriately. What is *signal*? To define this precisely is a difficult task. *Anything which carries information is a signal…* (italics added)

A signal is typically described as a *carrier* of message content. Thus, fire in the physical sphere creates smoke in the physical sphere that flows to the mental sphere to trigger the creation of an image or sense of fire. A signal is a carrier (itself) that includes content while traveling in a channel and may get loaded with noise. Here, *creation* in the FM model indicates the appearance in the communication process of a new thing (a carrier full of noise).

The basic features that differentiate carriers and content have fascinated researchers in the communication area. According to Reddy [28], "messages" are not contained in the signals; "The whole point of the system is that the alternatives (in Shannon's sense) themselves are not mobile, and cannot be sent, whereas the energy patterns, the 'signals' are mobile." Blackburn [29] insists that "messages are not mobile, while the signal is mobile." In FM, a thing is *conceptually* mobile since it flows. But conceptual flow is different from physical movement from one place to another. Flow is not necessarily a physical movement; for example, in the sphere of a *House*, the house "flows" from one owner to another. The paper will next use an FM diagram to illustrate the notion of signal through its content.

**Example**: Sang and Zhou [30] extend the BPMN platform to include specification of security requirements in a healthcare process. They demonstrate this through an example and show that BPMN standards cannot express the security requirements of such a system because of limitations in these standards; e.g., the Healthcare Server needs to execute an authentication function before it processes a Doctor's request. The example involves five components: (1) a Healthcare Device, a wearable device that senses a patient's vital functions such as blood pressure and heart rate, (2) a Healthcare Server, a cloud server that processes the patient's physical data, (3) a Display Device, (4) a Doctor, a medical expert who provides medical services, and (5) a Medical Device. Fig. 2 shows a partial view of the BPMN representation of the process.

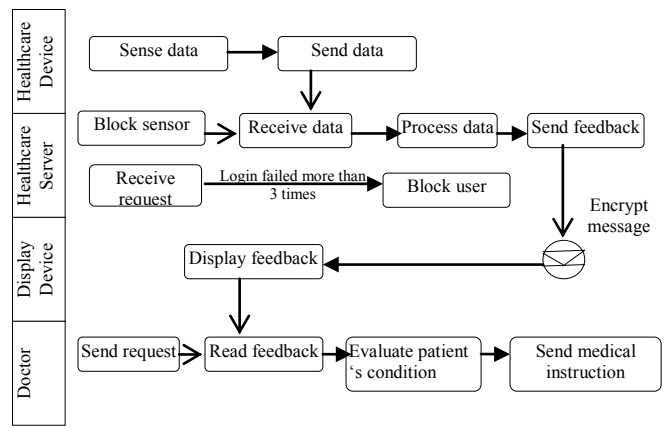

Fig. 2. BPMN representation (redrawn, partial from [30])

Fig. 3 shows the corresponding FM representation of the example as we understand it. In the figure, the sensor generates (1) data that flow to the server (2) to be processed (3) and generate feedback (4).

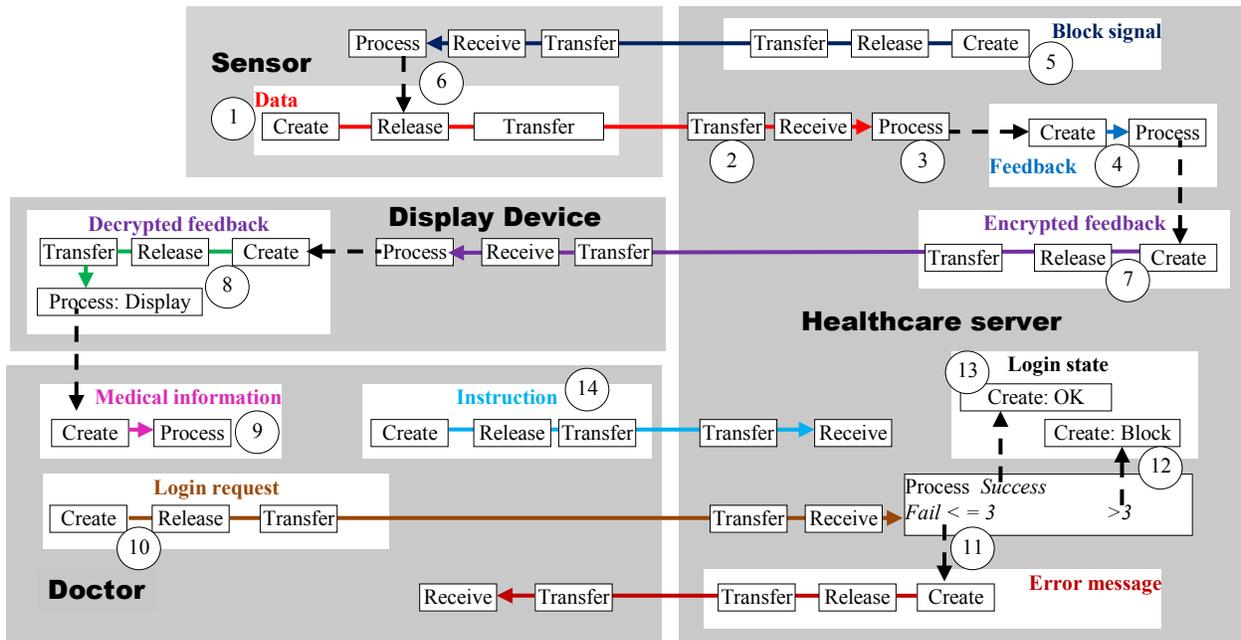

Fig. 3. FM representation of the example.

The server can create signals (5) to block the transmission of data from the sensor (6). The feedback is encrypted (7) and flows to the display device to be decrypted and displayed (8). The doctor reads the information and tries to login (10). The login attempt may fail up to three times (11). After that the login is blocked (12). If the login succeeds then the doctor inputs medical instructions to the system (14).

Fig. 3 is a static description. System behavior is modeled in terms of events. Here *behavior* involves the chronology of activities that can be identified by orchestrating their sequence in their interacting processes. In FM, *an event is a thing that can be created, processed, released, transferred, and received*. A *thing* becomes active in events. An event sphere includes at least the event itself, its time, and its region. For example, an event in this example is shown in Fig. 4: *Error message is sent to the doctor*. Accordingly, Fig. 5 shows selected events occurring in Fig. 3.

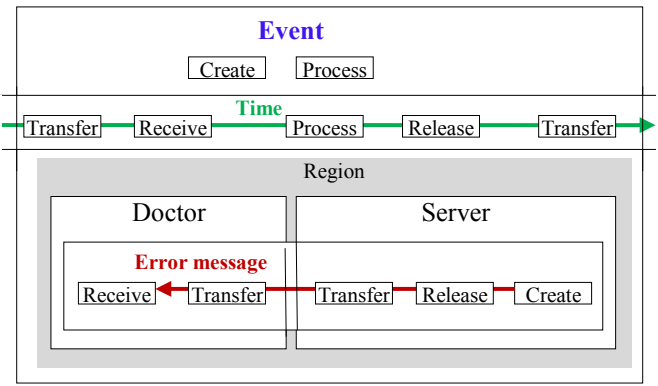

Fig. 4. The event *Error message is sent to the doctor*.

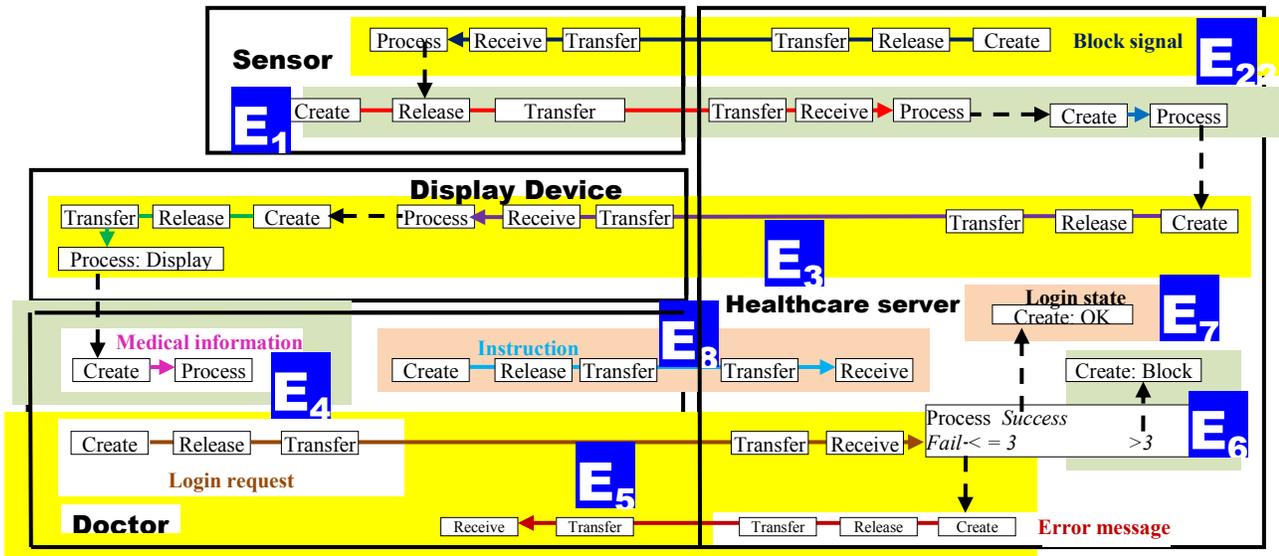

Fig. 5. Events in the healthcare process scenario.

To simplify the diagram we will omit the machines of time and of the event itself. The events are:

$E_1$: The sensor sends data to the server that are processed to create feedback.
$E_2$: The server blocks data from the sensor.
$E_3$: The feedback is encrypted and sent to the display device.
$E_4$: The doctor reads the displayed information.
$E_5$: The doctor tries to login and the login fails.
$E_6$: The login fails 3 times and is hence blocked.
$E_7$: The login succeeds.
$E_8$: The doctor sends medical instructions.

Accordingly, control of the system is defined as shown in Fig. 6.

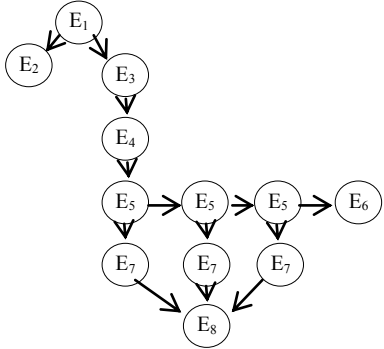

Fig. 6. Control sequence of the system

## IV. WHAT IS INFORMATION? WHAT ARE DATA?

*Data* are typically described as "raw information" or "things that have been given" [31]. In FM, "raw" refers to new-ness, a thing that has emerged or been created from outside the domain of the FM diagram. These raw data are different from manufactured data by processes in the FM diagram. The data have the possibility of *sliding* to become the content of a signal; thus the data are (in computer jargon) the sender and (part of) the "message" simultaneously, as seen in Fig 7. The raw data "ride" the signal to flow to another sphere (e.g., to be processed to trigger information). In physics, the sound of a bell is cut off in a vacuum because there are no signals (waves) to carry it when there is no surrounding air. Note that the purpose of this discussion is to apply it to persons and their PIIs.

A *raw* data machine (the flower in Fig. 7) lacks an agent of transfer; hence, it rides these signals. *Perceiving* a flower means *receiving* its signals of color, smell,… A "signal machine" is needed to carry it (e.g., rays of vision).

Consider another example of the four states of matter observable in everyday life: solid, liquid, gas, and plasma (see Wikipedia). Fig. 8 shows the occurrence of a signal as an *event*. An event can be described in terms of three components:

(i) Event region
(ii) Event time
(iii) Event itself as a thing.

Note that "processing" of the event itself refers to the occurrence of the event, and processing of time refers to time running its course.

This event must occur many times before the observing agent can reach the conclusion that there are four observable states. Accordingly, Fig. 9 shows this *repeated* experience of events until the recurrent information triggers the realization that there are only four observable states.

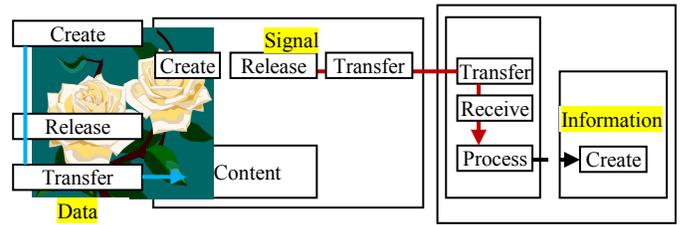

Fig. 7. Information is processed raw data.

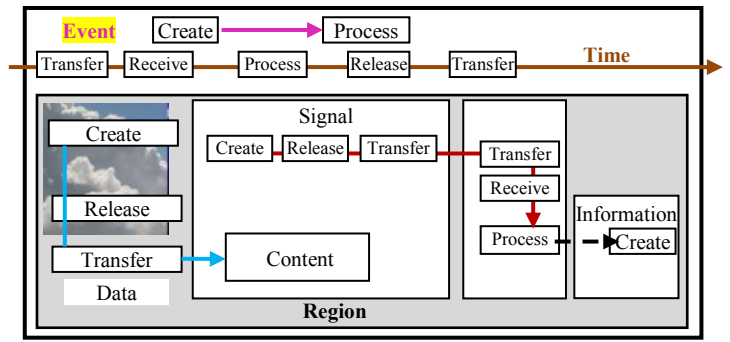

Fig. 8. The event *The creation of an information thing about a state of matter*.

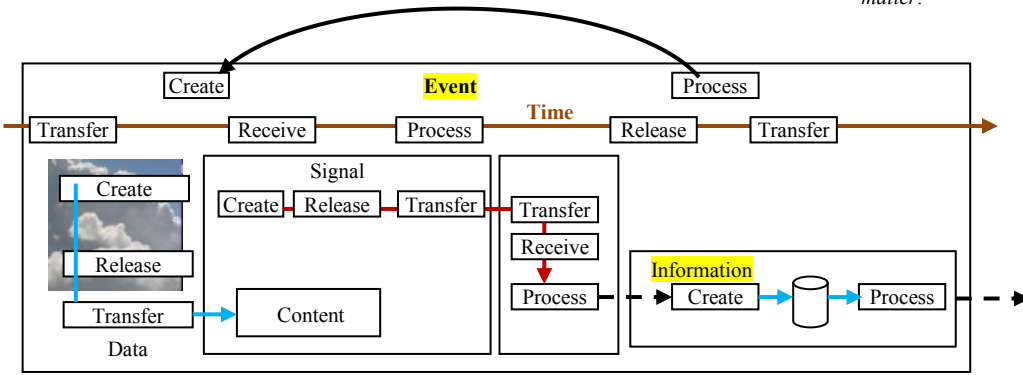

Fig. 9. Repeated event of *The creation of an information thing about a state of matter*.

Data can also be "manufactured," as is clear in Shannon's communication model. Fig. 10 shows how information about these states is generated from data directly and indirectly. First, the four observable states (1) "expose" themselves through signals (2) as information (3). Then, with sufficiently large events in which these phenomena occur, the informed agent can construct codes (4) in the form of data of signals (5) that flow to another informed agent (6). Note that in this case the data 00, 01, 10, and 11 are intentionally moved to fill the signal as its content (i.e., they do not *slide* to become content as in the case of flower). Certain pieces of information form identifiers, as described next.

## V. WHAT IS AN IDENTIFIER?

Meet Jean Blue, humanoid living in Centerville. Jean is a *real person, an identity*. Jean has many attributes, including gender, height, weight, preferred language, capabilities and disabilities, citizenship, voter registration, … [pieces of information]. Among these attributes are some *identifiers* … Identifiers are attributes whose values are specific and unique to an individual. [32]

A person's *identifier* can be constructed from things that identify (recognize) the person *uniquely*, e.g., characteristics and features. Identifiers are important for establishing the particularity or uniqueness of a person necessary for unique *identification* (i.e., recognition of a person). According to the Microsoft Word dictionary, identity is "the set of characteristics that somebody recognizes as belonging uniquely to himself or herself and constituting his or her individual personality for life." Grayson [33] expands this definition to include those characteristics about somebody that others recognize as well.

According to Grayson [33], "What we hear about identity (the noun) embodies more directly the notion of identify (the verb)… These notions are at best incompatible and, in the fullest understanding of identity, mutually exclusive." The definition of identity includes "belonging *uniquely* to . . . and constituting his or her individual personality . . . for life," thus "more than one identity for a given object means that object no longer has a unique identity." Put simply, if identity embodies identification and there are several methods of identifying a person, then a definition of identity that includes uniqueness seems contradictory.

We can use an identifier to refer to *recognizing a person uniquely*. According to Clarke [34], "Persona [identity] refers to the public personality that is presented to the world [and] supplemented, and to some extent even replaced, by the summation of the data available about an individual."

Problems occur in relation to the nature of data that materialize identifiers. What is "the data available about an individual"? Is the datum *John F. Kennedy is a very busy airport* about an individual named John F. Kennedy? Is the datum *John loves Alice* about John or Alice? We will use the term *identifier* to refer to things that identify (recognize) an individual *uniquely* in a specific sphere (context).

The Aristotelian entity is a single, specific existence (a particularity) in the world. In FM, as shown in Fig 11 (circles 1–3), an identifier of an entity can be its natural descriptors (e.g., 6 feet tall, brown eyes, male, blood type A, actions, etc.). Accordingly, an identifier is a thing that is processed to identify a (natural) person uniquely. Note the context in the figure related to PII in space, e.g., location and time. Consider the example of a privacy policy given by Finin et al. [35]:

> Do not allow my social network colleagues group (identity context) to take pictures of me (identity context) at parties (activity context) held on weekends (time context) at the beach house (location context).

Fig. 12 expresses diagrammatically the prohibited situation: *Social network colleagues group take pictures of me at parties held on weekends at the beach house.*

Consider the set of unique *identifiers* of persons. Ontologically, as mentioned, the Aristotelian entity/object is a single, specific existence (a particularity) in the world that comprises *natural descriptors* as communicated by *signals* These descriptors *exist* in the entity/object. Height and eye color, for example, exist as aspects of the existence of an entity.

Some descriptors form *identifiers*. A *natural identifier* is a set of natural descriptors that facilitate recognizing a person *uniquely*. We create an identifier (e.g., name) for a "specific" newborn baby (specific physical place and relationships). An identifier can also be created from the activities and actions of a person (circles 4 and 5 in Fig. 11).

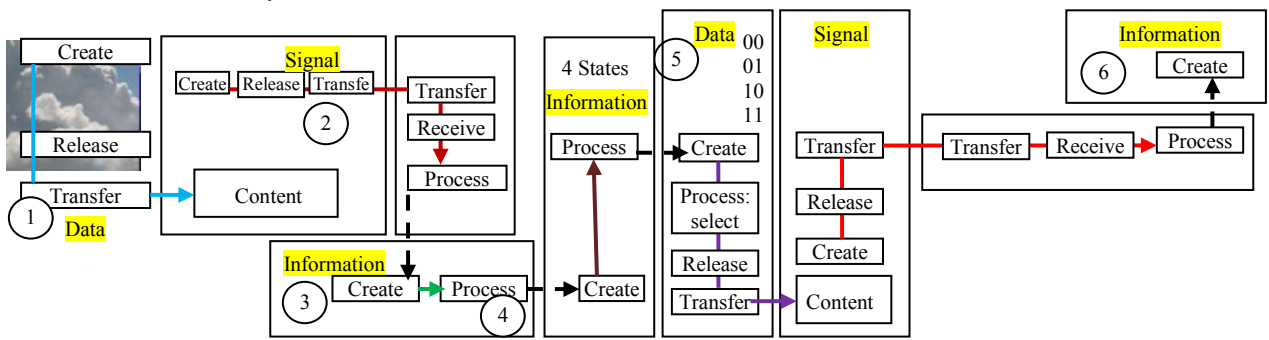

Fig. 10. Information coded as data.

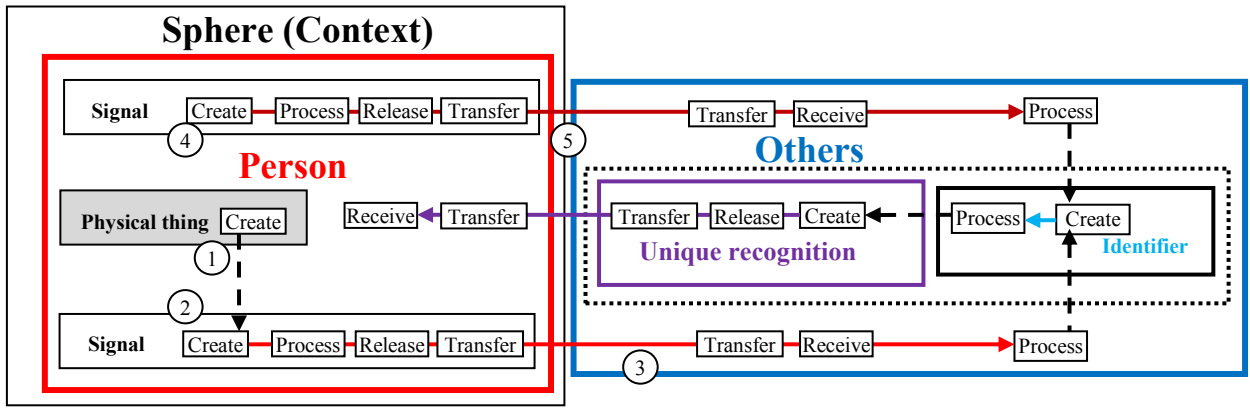

Fig. 11. An identifier is a thing created from data of a person as a physical thing or from data created by him/her that triggers unique recognition of that person within a sphere.

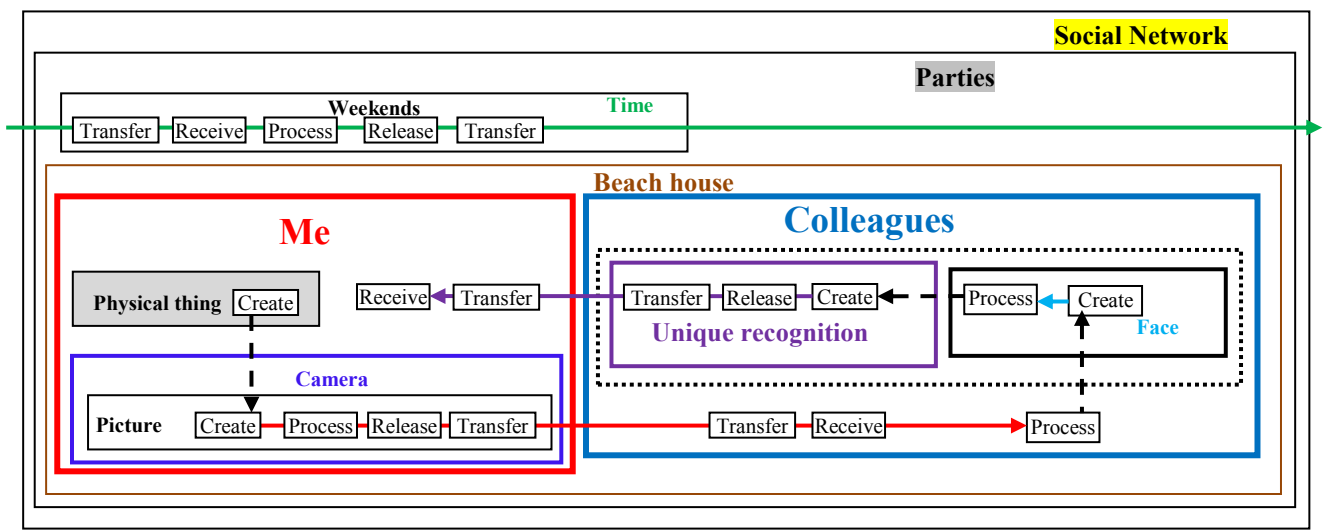

Fig. 12. The specification: *Social network colleagues group take pictures of me at parties held on weekends at the beach house.*

Note that an identifier is not necessarily sufficient to identify a person uniquely; we also need a recognition machine (dotted box in Fig. 11) that connects the identifier to the person. In reality, an "identifier" is insufficient to recognize a person, e.g., many people share the same name. This recognition implies knowing "who somebody is" or "the ability to get hold of them" as physical "bodies" [36]. "Simply to know a person's name is obviously not to know who that person is, even when the name in question is unique. ... We can also know who someone is without knowing their name" [36].

The dictionary definition of "identification" includes "act of identifying" as well as "evidence of identity." The "act" of identifying refers to pointing at or mapping to an individual. Similarly, "evidence" of identity refers to mapping this evidence to an individual. Typically, the "identity" itself is tied to physical existence. The identity of a "real" individual is "the individual's legal identity or physical 'meat space' location" [37], and "to identify the parties to a contract is to make it possible to hale them into court if they violate the contract. Identity, in other words, is employed as a means of access to a person's body" [36]. Thus, "identity" is something that distinguishes one "meat space" from another. This "something" is clearly a type of information. It is also "private" because it uniquely identifies this ontological space. Hence an identifier is the information aspect of the ontological space occupied by a human. Names, Social Security number, pictures, physical descriptions, fingerprints, and other identification devices are pointers to this "ontological space." We can recognize "identity" directly without using any of these pointers. When a witness "identifies" an offender from among other suspects in a police lineup, the witness recognizes this "ontological space" [36].

VI. WHAT IS PERSONAL IDENTIFIABLE INFORMATION, PII?

Information privacy "involves the establishment of rules governing the collection [in FM, *Receive*] and handling [in FM, *Process*] of *personal* [in FM, PII] data such as data in credit, medical, and government records. It is also known as "data protection" [38]. In the strict context of limiting privacy to matters involving information, the concept of privacy has been fused with PII protection [39]. In this context, PII denotes information about identifiable individuals in accessible form [40].

PII means any information concerning a natural person that, because of name, number, symbol, mark, or other identifier, can be used to identify that natural person [41]. It includes name or any identifiable number attached to it plus any other information triggered by signals such as address (location), telephone number, sex, race, religion, ethnic origin, sexual orientation, medical records, psychiatric history, blood type, genetic history, prescription profile, fingerprints, criminal record, credit rating, marital status, educational history, place of work, personal interests, favorite movies, lifestyle preferences, employment record, fiscal duties, insurance, ideological, political, or religious activities, commercial solvency, banking or saving accounts, real estate rental and ownership records.

Also, PII is "(t)hose facts, communications, or opinions which relate to the individual, and which it would be *reasonable to expect him to regard as intimate or sensitive* and therefore to want to withhold or at least to restrict their collection, use or circulation" [40] (italics added). The British Data Protection Act of 1984 defines PII ("personal data") as "information which relates to a *living* individual who can be identified from that information (or from that and other information in the possession of the data user), including any expression of opinion about the individual but not any indication of the intentions of the data user in respect of that individual … which is recorded in a form in which it can be processed by equipment operating automatically in response to instructions issued for that purpose" [42] (Italics added). The assumption here is that this PII is factual information (i.e., not libel, slander, or defamation). Jones [13] categorized six "senses" of PII (calling it personal data): information that is controlled or owned by or about us, directed toward us, sent by us, experienced by us, or relevant to us. The U.S. Department of Health & Human Services [43] defines PII in an IT system or online collection as information (1) that directly identifies an individual, or (2) by which an agency intends to identify specific individuals in conjunction with other data elements, i.e., indirect identification. The U.S. Department of Homeland Security (DHS) defines PII as "*Any information that permits the identity of an individual to be directly or indirectly inferred, including any information which is linked or linkable to that individual*" [44].

These are sample definitions of PII. In the context of FM, PII is defined as shown in Fig. 13. A single identifiable person is "the physical 'meat space' location" [37] and the identifier "is employed as a means of access to a person's body" [36].

Personal identifiable information (PII) is vital in today's privacy legislation, according to Schwartz and Solove [45]:

Personally identifiable information (PII) is one of the most central concepts in information privacy regulation. The scope of privacy laws typically turns on whether PII is involved. The basic assumption behind the applicable laws is that if PII is not involved, then there can be no privacy harm.

## VII. WHAT IS PRIVACY?

The world "private" derives from the Latin *privatus*, meaning "withdrawn from public life" or "deprived of office" [46]. The dictionary meaning of privacy includes the state of being private and undisturbed, freedom from intrusion or public attention, avoidance of publicity, limiting access, and the exclusion of others [47]. Privacy supports the conditions for a wide range of concepts including seclusion, retirement, solitude, isolation, reclusion, solitariness, reclusiveness, separation, monasticism, secretiveness, confidentiality, intimacy, anonymity, and to be left alone, do as we please, and control information about oneself. It is also an umbrella term that includes diverse contexts such as private places or territorial privacy, private facts or activities, private organizations, private issues, private interests, and privacy in the information context [48]. In general, privacy is also described as "the measure of the extent an individual is afforded the social and legal space to develop emotional, cognitive, spiritual and moral powers of an autonomous agent" [46]. It is "the interest that individuals have in sustaining a 'personal space', free from interference by other people and organizations" [49].

The notion of privacy as the *right to control "personal" information* has roots in the concept of individual *liberty*. Philosophically, liberty means freedom from some type of control. Liberty implies the ability to control one's own life in terms of work, religion, beliefs, and property, among other things. Historically, the *right* to control one's own *property* is a significant indicator of liberty. An owner can use, misuse, give away or dispose of his or her own property. Similarly, privacy is a personal thing "owned" by individuals, and they "control" it. Informational privacy is "the right to exercise some measure of control over information about oneself" [50].

In FM, we can view privacy on the basis of identifiers; in this case, privacy is *cutting off* sources of manufactured identifiers, as shown in Fig. 14. It is a restriction of flows of signals between a person and others. Fig. 14 is a version of Fig. 11, with the identifier machine deleted. Westin has defined privacy as the "claim of individuals, … to determine for themselves how, when, and to what extent information about them is communicated to others" [50].

It is common in the literature to define privacy as *Being in control of who can access information about the person*. This concept is represented in Fig. 15, where the release of data about a person is triggered by the person him or herself. Privacy may also be described as *Times when the person is completely alone, away from anyone else*, as shown in Fig. 16.

The point here is that the FM language is reasonably precise for expressing diverse conceptualizations of *what is privacy?* that can be related and analyzed in a unified framework.

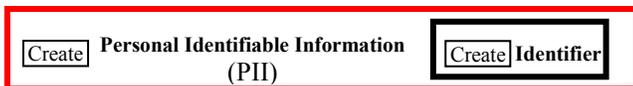

Fig. 13. Definition of PII

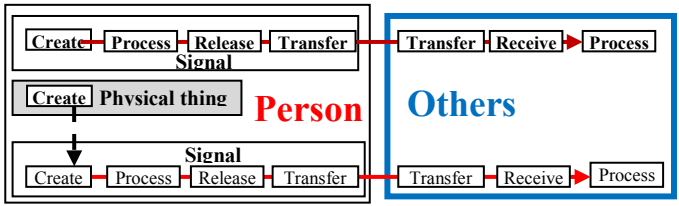

Fig. 14. *Privacy* is "cutting off sources" of manufactured identifiers

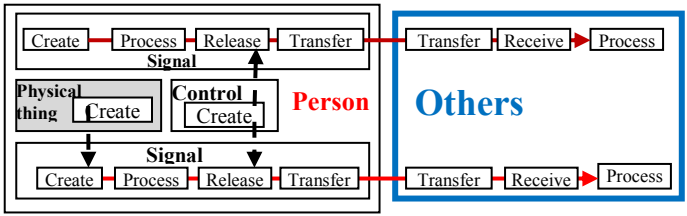

Fig. 15. *Privacy* is *Being in control of information about the person*.

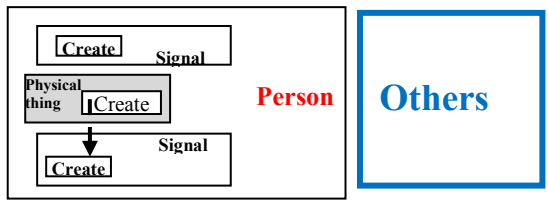

Fig. 16. *Privacy* is *Times when the person is completely alone, away from anyone else*.

## VIII. TYPES OF PII

In *linguistic* forms of information, we consider *assertion* a basic component. Language is the main vehicle that describes things and their machines in the domain of knowledge. In linguistic-based privacy, PII is an element that points uniquely to a single person-thing (type person). PII essentially "makes a person known," a potentially sharable entity that can be passed along in "further sharing." The classical treatment of assertion (judgment) such as PII divides it into two concepts: subject (referent) and predicate that form a logical relation; however, FM PII may or may not be a well-structured linguistic expression. The linguistic internal structure of any assertion is not the element of interest; rather it is its *referent*. *Newton is genius*, *Newton genius*, *genius Newton*, *Newton genius is*, *Newton is x*, *y Newton x*—are assertions as long as *Newton* is an identifier. Eventually, even a linguistic expression with one word such as *Newton* is a PII in which the non-referent part is empty.

PII is any information that has *referent*(s) of type natural persons. There are two types of personal information:
(1) Atomic PII (APII) is PII that has a single human referent.
(2) Compound PII (CPII) is PII that has more than one human referent. Fig. 17 shows a binary CPII. A CPII is reducible to a set of APIIs and a relationship, as is made clear in Fig. 17 For example, the statement *John and Mary are in love* can be privacy reducible to *John and someone are in love* and *Someone and Mary are in love*.

In logic (correspondence theory), *reference* is the relation of a word (logical name) to a *thing*. Every PII refers to its referents in the sense that it "leads to" him/her/them as distinguishable things in the world. This reference is based on his/her/their unique identifier(s).

A single referent does not necessarily imply a single occurrence of a referent. Thus, "*John* wounded *himself*" has one referent. *Referent* is a "formal semantics" notion [51] built on any linguistic structure such that its *extension* refers to an individual (human being).

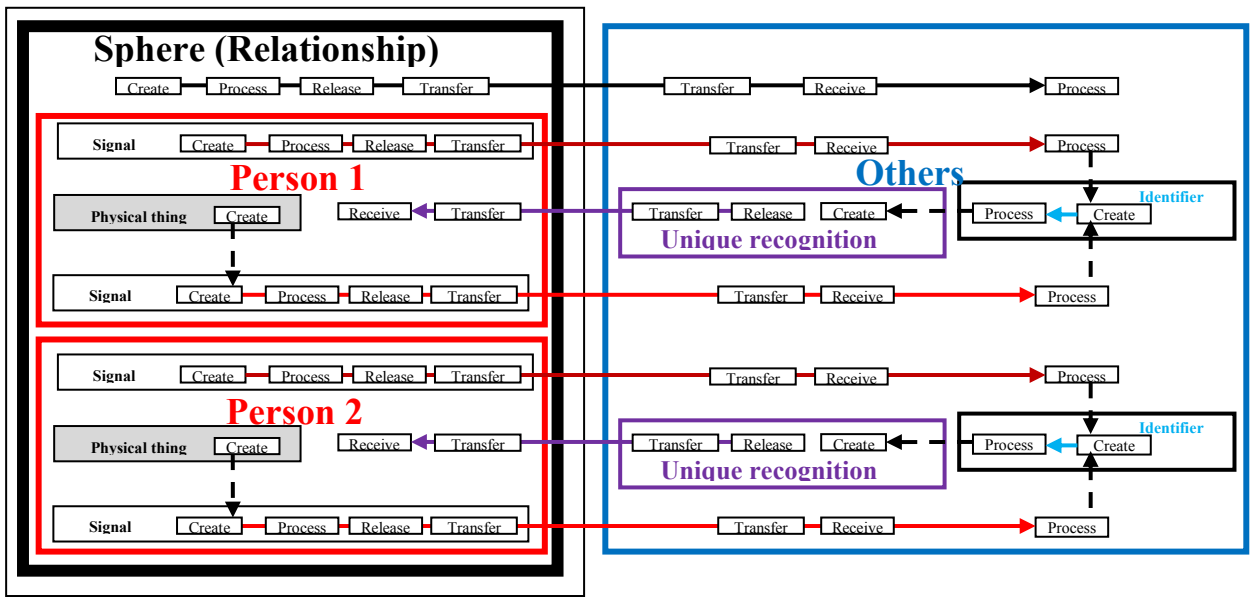

Fig. 17. Binary CPII

In logic, reference is the relation of a word (logical name) to a thing [52][53]. In PII, this *thing* is limited to human beings. In logic language, CPII is a predicate with more than one argument. Here the term "predicate" is used loosely since internal structure is immaterial. If we restrict FM to the language of first order logic, then its predicates are applied to logical names that refer to (natural) persons only. A piece of APII is a monadic predicate, whereas CPII is a many-place predicate. A three-place logical predicate such as *give(John, George, pen)* is a two-place predicate in FM since it includes only two individuals. In FM, it is assumed that every many-place predicate represents that many monadic predicates. *Loves(x1, x2)* represents *loves(x1)* and *being-loved(x2)*. Accordingly, *loves(x1, x2)* is private with respect to x1 because of *loves(x1)*, and it is private with respect to x2 because of *being-loved(x2)*. APII is the "source" of privacy. CPII is "private" because it embeds APII.

## IX. PROPRIETORSHIP OF PII

We call the relationship between PII and its referent *proprietorship*, such that the referent is the *proprietor*. The proprietorship of PII is conferred only to its proprietor. CPII is proprietary information of its referents: all donors of pieces of APII that are embedded in the compound PII.

Proprietorship is not Ownership. Historically, the rights to property were gradually legally extended to intangible possession such as processes of the mind, works of literature and art, good will, trade secrets, and trademarks [54]. In the past and in the present, private property has facilitated a means to protect individual privacy and freedom [55]; however, even in the nineteenth century it was argued that "the notion of privacy is altogether distinct from that of property" [56].

A proprietor of PII may or may not be its possessor and vice versa. Individuals can be proprietors or possessors of PII; however, non-individuals can be only possessors of PII. Every piece of APII is a proprietary datum of its referent. Proprietorship is a nontransferable right. It is an "inalienable right" in the sense that it is inherent in a human being. Others may have a "right'" to it through possessing or legally owning it but they are never its proprietor. Proprietorship of PII is different from the concept of copyright.

Copyright refers to the right of ownership, to exclude any other person from reproducing, preparing derivative works, distributing, performing, displaying, or using the work covered by copyright for a specific period of time [57]. In privacy the (moral) problem is more than "the improper acquisition and use of someone else's property, and ... the instrumental treatment of a human being, who is reduced to numbers and lifeless collections of information" [58]. It is also more than "the information being somehow embarrassing, shameful, ominous, threatening, unpopular or harmful." Intrusion on privacy occurs even "when the information is ... innocuous" [58]. "The source of the wrongness is not the consequences, nor any general maxim concerning personal privacy, but a lack of care and respect for the individual" [58]. Treating PII is equivalent to "treating human beings themselves" [58].

It is also important to notice the difference between *proprietorship* and *knowing* of PII. Knowing here is equivalent to possession of PII. APII of x is proprietary information of PII but it is not necessarily "known" by x (e.g., personal medical tests of employees). Possession-based "knowing" is not necessarily a cognitive concept. "Knowing" varies in scope; thus, at one time there may be a piece of APII "known" only by limited number of entities that then becomes "known" by more entities.

The concept of proprietorship is applied to CPII, which represents "shared proprietorship" but not necessarily shared possession or "knowing." Some or all proprietors of compound private information may not "know" the information.

## X. TRIVIAL PII

According to our definition of PII, every bit of information about a singly identified individual is his/her atomic PII. Clearly, much PII is trivial. *Newton has two hands*, *Newton is Newton*, *Newton is a human being*, etc. are all trivial bits of PII of Newton. Triviality here is the privacy counterpart of analytics in logic. Analytical assertions in logic are those assertions of which we can determine their truth without referring to the source. An assertion such as *All human beings are mortals* is true regardless of who says it. According to Kant, an analytical assertion is *a priori* and does not enlarge our knowledge. This does not mean that analytical assertions are insignificant; the opposite is true, in that all axioms of logic (e.g., principles of contradiction) are of this type. Similarly, trivial PII is privacy-insignificant. We will assume that PII is non-trivial.

The definition of PII implies embedding of identifiers. While identifiability is a strict measure of PII, sensitivity is a notion that is hard to pin down.

## XI. PII SENSITIVITY

Spiekermann and Cranor [1] introduce "an analysis of privacy sensitive processes" in order to understand "what user privacy perceptions and expectations exist and how they might be compromised by IT processes … to understand the level of privacy protection that is required." Accordingly, they claim:

> All information systems typically perform one or more of the following tasks: data *transfe*r, data *storage* and data *processing*. Each of these activities can raise privacy concerns. However, their impact on privacy varies depending on how they are performed, what type of data is involved, who uses the data and in which of the three spheres they occur. [Italics added]

FM introduces a more comprehensive view of these tasks. In general, the notion of sensitivity is a particularly difficult concept.

Defining PII as "information identifiable to the individual" does not mean that PII is "especially sensitive, private, or embarrassing. Rather, it describes a relationship between the information and a person, namely that the information—whether sensitive or trivial—is somehow identifiable to an individual" [59]. The *significance* of PII derives from its privacy value to a human being.

From an informational point of view, an individual is a bundle of his or her PII. PII comes into being not as an independent piece of information, but rather as a constitutive part of a particular human being [58]. PII ethics is concerned with the "moral consideration" of PII because PII's "well-being" is a manifestation of the proprietor's welfare [60].

There is a point that must be exceeded before beginning to consider PII sensitive. Social networks depend on the fact that individuals willingly publish their own PII, causing more dissemination of sensitive PII that compromises individuals' information privacy. This may indicate that PII sensitivity is an evolving notion that needs continuous evaluation. On the other hand, many Privacy-Enhancing Technologies (PETs) are being devised to help individuals protect their privacy [61], indicating the need for this notion.

The sensitivity of PII is a crucial factor in determining an individual's perception of privacy [62]. In many situations, sensitivity seems to depend on the context, and this cannot always be captured in a mere linguistic analysis; however, this does not exclude the possibility of "context-free" sensitivity (see [22]).

A typical definition of sensitivity of PII refers to the impact of handling (e.g., disclosing) of PII, as shown in Fig. 18.

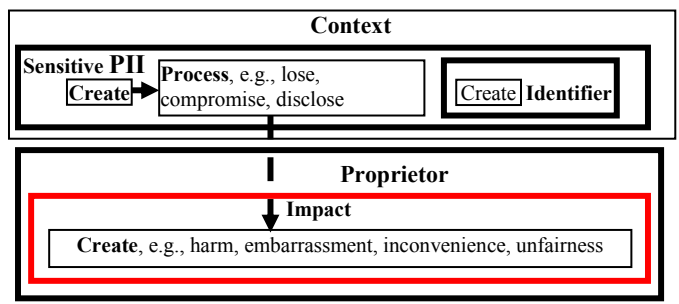

Fig. 18. Sensitive PII

## XII. MISINFORMATION

Consider the APII *John is honest*. Suppose that it is a true assertion. Does this imply that *John is dishonest*, which is false, is not PII? Clearly, this is not acceptable. Describing *John* as honest or dishonest is a privacy-related matter regardless whether the description is true or not. That is, "non-information" about an individual is also within the domain of his/her privacy.

## XIII. CONCLUSION

This paper has defined a fundamental notion of privacy: PII based on the notion of "things that flow." The resultant conceptual picture includes signals in communication and information and clarifies the sequence of ontological spaces and their relationship associated with these concepts. Clarifying these concepts is a beneficial contribution to the field of information privacy.

Further work can be directed toward developing a more elaborate model of types of privacy, especially in the area of sensitivity. Additional work includes *PII sharing* involving proprietors, possessors, and sharers (e.g., senders, receivers) of PII.